\documentclass[twocolumn,showpacs,superscriptaddress,prb,amsmath,amssymb,letterpaper]{revtex4}
\usepackage{times}
\usepackage{amsmath,bm,amsfonts}
\usepackage{dcolumn}
\usepackage{graphicx}
\usepackage{latexsym}

\begin{document}

\title{Valence Bond Solid Phases on Deformed Kagome Lattices: Application to Rb$_{2}$Cu$_{3}$SnF$_{12}$}

\author{Bohm-Jung \surname{Yang}}
\affiliation{Department of Physics, University of Toronto,
Toronto, Ontario M5S 1A7, Canada}

\author{Yong Baek \surname{Kim}}
\affiliation{Department of Physics, University of Toronto,
Toronto, Ontario M5S 1A7, Canada}
\affiliation{School of Physics,
Korea Institute for Advanced Study, Seoul 130-722, Korea}

\date{\today}

\begin{abstract}
Motivated by a recent experiment on Rb$_{2}$Cu$_{3}$SnF$_{12}$,
where spin-1/2 Cu$^{2+}$ moments reside on the layers of Kagome-like
lattices, we investigate quantum ground states of the antiferromagnetic Heisenberg model
on a series of deformed Kagome lattices. The deformation is characterized
by a weaker exchange coupling $\alpha J$ on certain lattice links
appropriate for Rb$_{2}$Cu$_{3}$SnF$_{12}$ with $\alpha=1$
corresponding to the ideal Kagome lattice. In particular, we study possible
valence bond solid phases using the perturbation theory around isolated dimer limits,
dimer series expansion, and self-consistent bond operator mean field theory.
It is shown that the valence bond solid phase with a 36-site unit cell of the ideal
Kagome lattice is quite sensitive to a small lattice distortion as the kind discovered
in Rb$_{2}$Cu$_{3}$SnF$_{12}$. As a result, we find that a more likely quantum
ground state in Rb$_{2}$Cu$_{3}$SnF$_{12}$
is the valence bond solid phase with a 12-site unit cell, where
six dimers form a pinwheel structure, leading to strong modification of the
elementary triplet and singlet excitation spectra in the deformed
Kagome lattices.
\end{abstract}


\maketitle

\section{\label{sec:intro} Introduction}

Lattice distortions are abundant in frustrated magnets as it is an efficient way to relieve
the frustration and lift the macroscopic degeneracy of the classical ground states.
This has been, for example, an obstacle to synthesize and study materials that
can be described by the quantum Heisenberg model with spin-1/2 moments
on ``ideal" lattice structures with isotropic spin exchange couplings.
The interest on the ``ideal" models stems from various theoretical suggestions for
novel emergent quantum paramagnetic phases such as quantum spin liquid\cite{Misguich_book,Fa_PSG,Hastings,Ran,sachdev_kagome, Ryu}
and valence bond solid (VBS)\cite{Huse,Huse2,Marston,Nikolic}
phases. On the other hand, those deformations may lead to different kind of emergent
quantum ground states and understanding precisely how these phases are
connected to the quantum ground states in the ``ideal" limit would help us
understand more global picture of the quantum frustrated magnets.\cite{Lawler, Fa}

One of the most studied frustrated magnets is the quantum spin-1/2 antiferromagnetic Heisenberg model
on the Kagome lattice. There exist several materials that are candidates for the realization
of this system.\cite{Exp,Exp_herbert1,Exp_herbert2,Exp_herbert3,Exp_mag1,Exp_mag2,Exp_vesign,Exp_volbo}
It has been, however, found that many materials undergo lattice deformations
and quite often this leads to magnetically ordered phases at sufficiently low temperatures.\cite{Exp_mag1,Exp_mag2}
So far, there are a small number of examples of two-dimensional Kagome or Kagome-like materials with
spin-1/2 moments that do not show any magnetic ordering down to very low temperatures.\cite{Exp,Exp_herbert1,Exp_herbert2,Exp_herbert3,Exp_vesign,Exp_volbo}
The Herbertsmithite, ZnCu$_{3}$(OH)$_{6}$Cl$_{2}$,\cite{Exp_herbert1,Exp_herbert2,Exp_herbert3}
is known for the ``ideal" Kagome lattice
structure, but it suffers from inter-site mixing between Cu$^{2+}$ and Zn$^{2+}$, and makes the
interpretation of the susceptibility data difficult. In the Volborthite,
Cu$_{3}$V$_{2}$O$_{7}$(OH)$_{2}$$\cdot$2H$_{2}$O,\cite{Exp_volbo} the Kagome structure is deformed
in an orthorhombic fashion, leading to two inequivalent spin exchange couplings between
Cu$^{2+}$ moments. In both cases, the specific heat at low temperatures reveals the absence
of (or a very small) excitation/spin gap (down to 50mK in Herbertsmithite, for example),
leading to various suggestions for exotic quantum ground states.\cite{Ran, Ryu, Huse}
More recently, it is found that Cu$^{2+}$ moments in Rb$_{2}$Cu$_{3}$SnF$_{12}$ form
the layers of a deformed Kagome lattice and do not show any magnetic order at low temperatures.\cite{Exp}
The susceptibility data clearly shows a spin excitation gap (in contrast to the previous two materials)
which is of the order of 20K while the average strength of the exchange coupling is estimated to
be around 200K, which leads to the expectation that the ground state may be a VBS state.

\begin{figure}[t]
\centering
\includegraphics[width=8 cm]{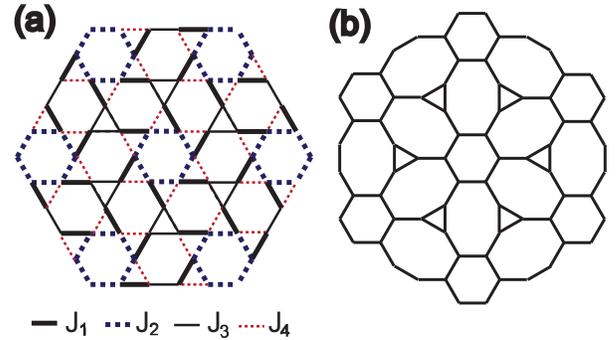}
\caption{(Color online)
(a) The deformed Kagome lattice with four different
neighboring exchange interactions, which may be realized in Rb$_{2}$Cu$_{3}$SnF$_{12}$.
Here all the exchange coupling constants are positive and satisfy
$J_{1} > J_{2} > J_{3} \gg J_{4}$. In this work, we consider the case of
$J_{1} = J_{2} = J_{3} = J > J_{4}$.
(b) The J$_{4}$-depleted Kagome lattice that is topologically
equivalent to the deformed Kagome lattice when $J_{1}=J_{2}=J_{3}=J$
but $J_{4}=0$.
} \label{fig:LatticeStructure}
\end{figure}

In this paper, we investigate the quantum ground states of the spin-1/2 antiferromagnetic Heisenberg model on
a series of deformed Kagome lattices that may be relevant to Rb$_{2}$Cu$_{3}$SnF$_{12}$.
Each deformed Kagome layer can be described by four possibly different nearest-neighbor exchange
couplings, labeled by $J_{1}$, $J_{2}$, $J_{3}$, and $J_{4}$ in decreasing order in magnitude
as inferred from different Cu$^{2+}$ -- F$^-$ -- Cu$^{2+}$ bond angles (See Fig.~\ref{fig:LatticeStructure}(a)).
In Rb$_{2}$Cu$_{3}$SnF$_{12}$, $J_{1}$, $J_{2}$, and $J_{3}$ are similar in magnitude
while the magnitude of $J_{4}$ is about a half of the others.\cite{Exp} Hence it is reasonable to
start from a model where $J_{1}$ = $J_{2}$ = $J_{3}$ = $J \gg J_4$.
Here we study the evolution of the quantum ground states as a function of
$\alpha = J_4/J \le 1$, where $\alpha=1$ corresponds to the ideal Kagome lattice.

Using perturbative arguments starting from isolated dimer limits, we find that there exist two
competing VBS phases on these lattices; one with a 12-site unit cell (VBS-12, shown in Fig.\ref{fig:VBSStructure}(a))
and another with a 36-site unit cell (VBS-36, shown in Fig.\ref{fig:VBSStructure}(b)).
The VBS-36 phase is smoothly connected to the VBS phase with a
large unit cell on the ideal Kagome lattice, which has been suggested as a strong contender for
the quantum ground state of the spin-1/2 antiferromagnetic Heisenberg model.\cite{Huse,Marston,Nikolic}
The perturbation theory clearly shows that the VBS-12 is the lower energy
ground state when $\alpha=0$, namely on the $J_4$-depleted lattice. On the other
hand, the VBS-36 becomes the ground state when $\alpha=1$, consistent with the
previous studies on the ideal Kagome lattice.
We study the relative stability of these phases for an arbitrary deformation parameter $\alpha$
using the perturbation theory, dimer series expansion, and bond operator mean field theory.
Thus our work can also test the stability of the VBS phase with a 36-site unit cell on the ideal Kagome lattice
against the deformation from the ideal Kagome lattice structure.

It is found that the VBS-36 phase on the ideal Kagome lattice is quite sensitive to the deformation
and only 3$\%$ reduction in the magnitude of $J_4$ is enough to induce the instability of the
VBS-36 state of the ideal Kagome lattice. Thus we propose that the ground state for $\alpha < 0.97$ is
the VBS-12 state. Note that the VBS-12 state is clearly the ground state of $J_1$-only model or
when $J_2, J_3 \ll J_1$. It is remarkable that the VBS-12 state is stable even when the strength
of $J_2$ and $J_3$ are equal to $J_1$.
Thus we conclude that the VBS phase discovered in Rb$_{2}$Cu$_{3}$SnF$_{12}$
is most likely the VBS-12 state. In order to provide a future experimental test for the VBS-12
state, we compute the triplet dispersion spectra as shown
in Fig.\ref{fig:Dispersion}.

The rest of the paper is organized as follows.
In Sec.\ref{sec:candidate}, we use perturbative arguments
about isolated dimer limits to show that there exist two competing
VBS phases - the VBS-12 and VBS-36 - on the $J_4$-depleted lattice.
In Sec.\ref{sec:groundproperties}, the dimer series
expansion and the bond operator mean field theory
are developed to investigate the relative stability of the
two candidate VBS phases for an arbitrary strength of $J_4/J$.
The triplet dispersion spectra
is also discussed in detail for the VBS-12 state. Finally,
in Sec.\ref{sec:discussion}, we discuss the implications of
our results to theory and experiments.

\section{\label{sec:candidate} Valence Bond Solid Ground States on the J$_{4}$-Depleted Kagome Lattice}

\subsection{\label{sec:secondorder} Second Order Perturbation and the VBS-12 State}

We can obtain valuable information about the spatial ordering
pattern of singlet dimers by applying the perturbation
theory around an isolated dimer limit.\cite{perturbation}
To perform the calculation,
we divide the Hamiltonian into two parts, {\it i.e.} $H=H_{0}+ \lambda V$.
Here $H_{0}$ describes the Heisenberg spin
exchange interaction between two nearest-neighbor spins constituting
a singlet dimer. The inter-dimer
interactions are contained in $V$ which is treated as a perturbation.
The ground state of the unperturbed Hamiltonian $H_{0}$, {\it i.e.}
$|\Psi_{0} \rangle$ can be taken as the direct product of a set of isolated dimer singlets
lying on the $J_{4}$-depleted Kagome lattice.

\begin{figure}[t]
\centering
\includegraphics[width=7 cm]{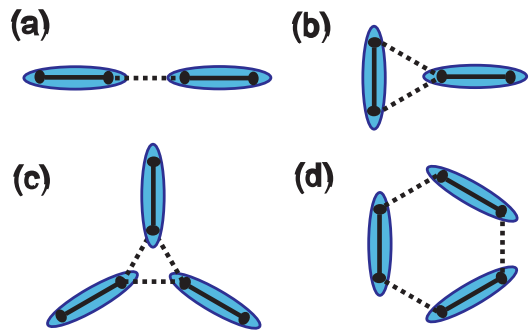}
\caption{(Color online)
Local configurations of neighboring dimers, which contribute to the lowest
order perturbation theory. Here an ellipse indicates a dimer singlet.
A solid line shows the intradimer coupling included in $H_{0}$
while a dotted line indicates the interdimer
coupling contained in the perturbation, $V$.
(a) Neighboring dimers are lying in parallel. There are
energy gains $\Delta \epsilon^{(2)} = - \frac{3}{32} J \lambda^2$ in the second
order and $\Delta \epsilon^{(3)}_{a} = - \frac{3}{128} J \lambda^3$ in the third
order.
(b) Neighboring dimers are lying in a perpendicular geometry.
There is no energy gain in the perturbation theory (explained in the text).
Here the triangle carrying a dimer is called a {\it filled} triangle.
(c) An {\it empty} triangle. There is no energy gain in the third order perturbation theory.
(d) A {\it perfect} hexagon. There is an additional energy gain
$\Delta \epsilon^{(3)}_{b} = - \frac{9}{128} J \lambda^3$ in the third
order perturbation theory.
} \label{fig:SecondOrder}
\end{figure}

Let us start from two possible local configurations of two neighboring dimer
singlets as shown in Fig.~\ref{fig:SecondOrder}(a) and (b). Note that
a pair of neighboring dimers can be aligned in parallel, connected by a single
unoccupied link (Fig.~\ref{fig:SecondOrder}(a)) or they can be
aligned in a perpendicular direction if one end of a dimer is simultaneously
connected to both spins of its neighboring dimer (Fig.~\ref{fig:SecondOrder}(b)).
The important point is that the ground state energy can be lowered through
second order perturbation processes only when the two dimers are lying in parallel.
When two dimers are lying in a perpendicular direction, $V$ $|\Psi_{0} \rangle$=0
because of the odd parity of the singlet state with respect to the reflection and there
is no energy shift due to the second order processes. The energy gain from each configuration
where two neighboring dimers are lying in parallel is $\Delta \epsilon^{(2)}=-\frac{3}{32}J \lambda^2$.

In order to determine the global structure of the VBS phases, however,
we need to take into account all the geometric constraints for possible
VBS patterns. Note that the basic building blocks of the $J_{4}$-depleted Kagome lattice
are triangles, hexagons and bridge links which connect each triangle
to its neighboring hexagons (shown in Fig.~\ref{fig:LatticeStructure}(b)).
One can show that the number of triangles, $N_{\triangle}$, is given by
$N_{\triangle} = \frac{N}{6}$, where $N$ is the number of spins (or lattice sites)
on the lattice. Similarly, the number of hexagons, $N_{\rm Hexagon}$,
and the number of bridge links, $N_{\rm Bridge}$, satisfy
$N_{\rm Hexagon} = \frac{N}{12}$ and $N_{\rm Bridge} = \frac{N}{2}$, respectively.
The total number of links, $L_{\rm Link}$, on the lattice is then given by
\begin{equation} \label{eq:link}
N_{\rm Link} = 3 N_{\triangle} + 6 N_{\rm Hexagon} + N_{\rm Bridge} = \frac{3}{2} N.
\nonumber
\end{equation}
Since the number of dimers, $N_{\rm Dimer}$, is equal to $\frac{N}{2}$ and each dimer
is lying on a single link, the total number of the unoccupied links is
$N_{\rm Link}-N_{\rm Dimer}=\frac{3}{2}N- \frac{1}{2}N = N$. The unoccupied
links connecting neighboring dimers in various local dimer configurations
are shown as the dotted lines in Fig.~\ref{fig:SecondOrder}

\begin{figure}[t]
\centering
\includegraphics[width=8 cm]{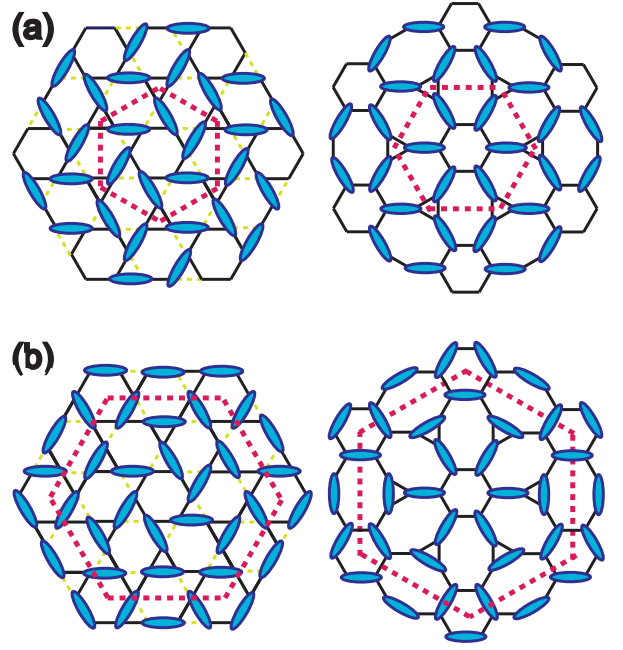}
\caption{(Color online)
Two candidate valence bond solid configurations on the
deformed Kagome lattice. The figure on the left-hand side is
the configuration on the original deformed Kagome lattice and
the picture on the righthand side represents a topologically equivalent
lattice of the $J_4$-depleted lattice.
The light dotted (yellow) lines on the lefthand-side pictures show the $J_4$ links.
(a) VBS-12: A valence bond solid with a 12-site unit cell.
(b) VBS-36: A valence bond solid with a 36-site unit cell.
Here the thick dotted (pink) lines indicate the unit cells.
} \label{fig:VBSStructure}
\end{figure}

According to the result of the second order perturbation theory for
two neighboring dimers, when one of the links of a triangle is occupied by
a dimer singlet (we call such a triangle as a {\it filled} triangle),
the other two links would generate no energy gain from the second order
processes because these two unoccupied links are simultaneously connected to one end of
a nearby dimer, generating a `perpendicular' arrangement between
the filled link of a triangle and a nearby dimer.
Hence if we denote the number of the
{\it filled} triangles as $N_{{\rm filled} \triangle}$, the number of
the unoccupied links that connect two nearby dimers in parallel configuration
is $N_{\rm Link}-N_{\rm Dimer}- 2 N_{{\rm filled} \triangle}$.
Therefore, the total energy gain per
spin from the second order processes is given by
\begin{align}\label{eq:E-secondorder}
\Delta E^{(2)} =&\Delta \epsilon^{(2)} \Big\{\frac{N_{\rm Link}-N_{\rm Dimer}- 2 N_{{\rm filled} \triangle}}{N}\Big\}
\nonumber\\
=&\Delta \epsilon^{(2)}\Big\{1-2\frac{N_{{\rm filled} \triangle}}{N}\Big\}.
\end{align}
Note that the condition $N_{{\rm filled} \triangle}$=0 uniquely determines
the ground state VBS configuration which is shown in Fig.~\ref{fig:VBSStructure}(a).
Here the unit cell consists of six dimer singlets which lie in a pinwheel structure
and we call this VBS state as the VBS-12 state.
In the VBS-12, every triangle is an {\it empty} triangle, that is,
no link of a triangle is occupied by a dimer singlet.
This is the dimer configuration one would obtain
when only $J_1$ is present or $J_2, J_3 \ll J_1$ (See Fig.~\ref{fig:LatticeStructure}(a)).
It is interesting to note that the VBS-12 state is stable all the way to the limit where
$J_2$ and $J_3$ become comparable to $J_1$.

\subsection{\label{sec:thirdorder} Third Order Perturbation and the VBS-36 State}

Now we extend the perturbation theory to the third order in $\lambda$.
Let us first consider two neighboring dimers lying in parallel
as shown in Fig.\ref{fig:SecondOrder}(a). In addition to
the energy gain from the second order perturbation, $\Delta \epsilon^{(2)}$,
this local configuration allows a third order process which results
in the energy gain, $\Delta \epsilon^{(3)}_{a}=-\frac{3}{128} J \lambda^3$.
On the other hand, the dimer configuration in Fig.\ref{fig:SecondOrder}(b)
does not allow any third order process. In fact, this configuration is locally
an eigenstate of the Hamiltonian and does not produce any virtual state in
any order of the perturbation theory.
Similarly, the unoccupied links in Fig.\ref{fig:SecondOrder}(c)
making an {\it empty} triangle, would not generate any energy gain in
the third order, either.
Therefore when we calculate the third order energy gain
coming from various local configurations,
we have to remember that the unoccupied links
on every {\it filled} and {\it empty} triangles would not
produce any third order energy shift.

Fig.\ref{fig:SecondOrder}(d) shows another configuration in which
a new third order process can be generated. This
structure, which is commonly known as a {\it perfect} hexagon,
consists of three neighboring dimers lying on a hexagon.
Through the resonating third order processes around
a {\it perfect} hexagon,
the ground state energy can be lowered by
$\Delta \epsilon^{(3)}_{b}=-\frac{9}{128}J \lambda^3$.

Based on these results,
the total energy gain per spin by the third order processes
can be summarized as follows.
\begin{align}\label{eq:E-thirdorder}
\Delta E^{(3)} =&\Delta \epsilon^{(3)}_{a} \Big\{\frac{N_{\rm Link}-N_{\rm Dimer}-
2 N_{{\rm filled} \triangle}- 3 N_{{\rm empty} \triangle}}{N}\Big\}
\nonumber\\
&\qquad+\quad \Delta \epsilon^{(3)}_{b} \frac{N_{\rm perfect}}{N}
\nonumber\\
=&\Delta \epsilon^{(3)}_{a}\Big\{\frac{1}{2}+\frac{N_{{\rm filled} \triangle}}{N}\Big\}
+\Delta \epsilon^{(3)}_{b} \frac{N_{\rm perfect}}{N},
\end{align}
where $N_{\rm perfect}$ represents the number of perfect hexagons
and we use the fact that the number of triangles,
$N_{\triangle}=N/6$, satisfies
$N_{\triangle}=N_{{\rm empty} \triangle}+N_{{\rm filled} \triangle}$.

Therefore if we can maximize the number of the filled triangles and
perfect hexagons, we can get the largest energy gain from the
third order processes. The VBS-36 state depicted in
Fig.\ref{fig:VBSStructure}(b) satisfies this requirement, that is,
$N_{{\rm filled} \triangle}=N_{\triangle}$ and
$N_{\rm perfect}=\frac{2}{3} N_{\rm Hexagon}$.
Note that the three neighboring hexagons lying in a triangular structure
cannot be `perfect' at the same time for any kind of dimer coverings.
Interestingly, this VBS-36 phase
is adiabatically connected to the 36-site-unit-cell VBS ground state
of the ideal Kagome lattice \cite{Huse, Nikolic} in the sense that
as we increase the magnitude of $J_{4}/J$ from zero to one,
the VBS-36 phase evolves smoothly to the 36-site-unit-cell
VBS ground state of the ideal Kagome lattice.

In the case of the $J_4$-depleted lattice, the VBS ground state is already
uniquely determined by the second order perturbation theory.
That is, the VBS-12 phase is the ground state and the VBS-36
phase is a closely competing phase.
However, it is not clear which phase is more stable when
$J_{4}$ has finite magnitude.
In the case of the ideal Kagome lattice,
all the dimer coverings are degenerate up to the second
order in $\lambda$ and the degeneracy can only be
lifted by the third order perturbation processes
as emphasized in Ref.\onlinecite{Huse}.
This is because the numbers of the {\rm filled} and {\rm empty} triangles
on the ideal Kagome lattice are fixed for any dimer covering
due to the spatial geometry of the ideal Kagome lattice (See Ref.\onlinecite{Nikolic}).
This is, in fact, the reason why the 36-site-unit-cell VBS is the most stable VBS state on
the ideal Kagome lattice.
Therefore, we may expect that the change in magnitude of $J_{4}$
would affect the relative importance between the second order
and the third order processes. In the next section, we
investigate the effect of finite $J_{4}$ on the ground state
properties of the deformed Kagome lattice.

\section{\label{sec:groundproperties} Valence Bond Solid Phases of the Deformed Kagome Antiferromagnet :
the Effect of Finite $J_{4}$}

\subsection{\label{sec:SeriesExpansion} Dimer Series Expansion Study}

In order to refine the ground state energy estimation beyond the third order
perturbation theory and confirm the results of
the previous lowest order expansions,
we develop a systematic scheme to compute the perturbation series using
the linked cluster expansion method. In this approach, the ground state energy can be expressed
as the sum of the contributions from a sequence of finite clusters \cite{SeriesBook, SeriesBook2}.
To carry out the series expansion, all the occupied links that make up the dimer covering
are given an interaction strength $J$ and all the other unoccupied links are given a strength $\lambda J$.
We apply the series expansion method to compute the
energy of the two competing VBS phases on the $J_4$-depleted lattice, $J_4=0$,
and the ideal Kagome lattice limit, $J_4/J=1$, up to the fifth order in $\lambda$.

The ground state energy per site of an infinite lattice can be written
as a sum over cluster terms of the form, $E_{0}/N=\sum_{g}c(g)\varepsilon(g)$.
Here $c(g)$ is the so-called lattice constant for the cluster $g$,
which is the number of embeddings of the cluster $g$ per lattice site.
$\varepsilon(g)$ is the reduced energy of the cluster $g$, which
is defined recursively as $\varepsilon(g)=E(g)-\sum_{g'}c(g'/g)\varepsilon(g')$.
Here $E(g)$ is the ground state energy of the cluster $g$ and
the sum is over all subclusters $g'$ of $g$.
Since the reduced energy $\varepsilon(g)$ is equal to the sum of the contributions from
all connected diagrams ({\it i.e.}~the patterns of connected dimers) that span the cluster $g$,
such diagrams only contribute beyond a certain minimum order in $\lambda$, proportional to
the size of the cluster. Thus, to obtain a perturbation series that is
exact up to a given order, it is sufficient to sum up the reduced energies
of all the clusters up to the corresponding size. This gives us a systematic
procedure for carrying out the expansion to higher orders.\cite{SeriesBook}


Another appealing aspect of applying the linked cluster expansion method is that
we can considerably reduce the complexity of the calculation by using the freedom
in choosing clusters. \cite{Rigol1,Rigol2}
In a recent work of the dimer series expansion study on the ideal Kagome lattice\cite{Huse},
the unoccupied links on each triangle are grouped together, which results in significant
simplification of the computation and rapid convergence of the series for the ground state energy.
In this scheme, for any dimer covering on the ideal Kagome lattice,
every {\it empty} and {\it filled} triangles
constitute the elementary clusters. All the other
clusters can be constructed by putting the elementary
clusters together appropriately. Based on this specially designed
clusters, it was shown that the series for the ground state energy
can be determined up to the fifth order in $\lambda$
using only five different graphs (or diagrams).\cite{Huse}
Summing up the terms in the series, the ground state energy
per spin of the VBS-36 phase on the ideal Kagome lattice is estimated
to be $-0.432 J$. For comparison, we compute the ground state
energy of the VBS-12 phase on the ideal Kagome lattice.
The same graphs can be used for the calculation and the summation
of the terms in the series up to the fifth order in $\lambda$
leads to the energy of $-0.424 J$. Note that even though
the VBS phase with a 36-site unit cell is the ground
state of the ideal Kagome lattice, the
energy difference between the two VBS phases is quite small.


\begin{figure}[t]
\centering
\includegraphics[width=8 cm]{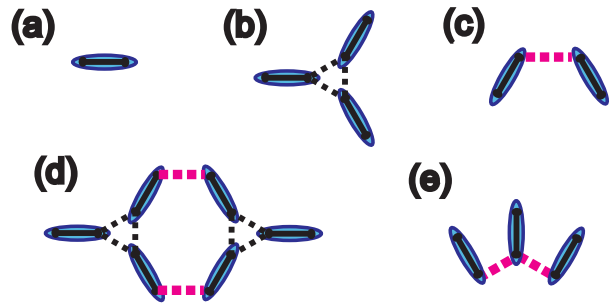}
\caption{(Color online)
Topologies of the graphs (diagrams) relevant to the VBS-12 phase
on the $J_{4}$-depleted Kagome lattice that contribute to the
ground state energy up to the fifth order in $\lambda$.
The thick dotted (pink) lines are the weak links on the hexagons while the thin
dotted (black) links represent the weak links on the empty triangles.
} \label{fig:VBS12-4th}
\end{figure}

Applying similar strategies to the $J_{4}$-depleted Kagome lattice,
we compute the series for the ground state energy of
the two competing VBS phases.
We first consider the VBS-12 phase
on the $J_{4}$-depleted Kagome lattice.
Here we group the three weak links on an empty triangle together and
distinguish them from the other weak links on the hexagons.
In this scheme, only five different graphs (shown in Fig.\ref{fig:VBS12-4th})
contribute to the ground state energy up to the fifth order in $\lambda$.
The smallest cluster in Fig.\ref{fig:VBS12-4th}(a)
is made of a single dimer and has
the reduced energy, $\varepsilon_{0, a}=-{3 \over 4} J$.
Fig.\ref{fig:VBS12-4th}(b) and (c) show two clusters whose
contributions start from the second order in $\lambda$.
The reduced energies of the corresponding clusters, {\it i.e.},  $\varepsilon_{2, b}$ and $\varepsilon_{2, c}$
respectively, can be obtained after subtracting the subcluster energy $\varepsilon_{0, a}$
appropriately and are found to be
\begin{align}\label{eq:second reduced energy}
\frac{\varepsilon_{2, b}}{J} &= - \frac{9}{32} \lambda^{2}
+ 0.03955078 \lambda^{4} - 0.01112366 \lambda^{6} + O(\lambda^{8}),
\nonumber\\
\frac{\varepsilon_{2, c}}{J} &= - \frac{3}{32} \lambda^{2}
-\frac{3}{128} \lambda^{3} - \frac{3}{2048} \lambda^{4} + \frac{15}{8192} \lambda^{5} + O(\lambda^{6}).
\nonumber
\end{align}
Finally, the resonance processes coming from the remaining two graphs
in Fig.\ref{fig:VBS12-4th}(d) and (e) can lower the ground state energy
through the fourth order processes. The corresponding reduced energies of the
two clusters, {\it i.e.}, $\varepsilon_{4, d}$ and $\varepsilon_{4, e}$
can be obtained by the same procedure used earlier and using the fact that
the subclusters are made of the graphs in Fig.\ref{fig:VBS12-4th} (a), (b) and (c).
Since the reduced energies of all the other clusters that are larger in size than those
in Fig.\ref{fig:VBS12-4th}(d) and (e), begin with O($\lambda^{6}$) terms, the ground
state energy can be determined exactly at least up to the fifth order in $\lambda$
with those five graphs shown in Fig.\ref{fig:VBS12-4th}.
After we determine the lattice constants of all the clusters, the ground state energy for the
infinite lattice is found to be
\begin{equation}\label{eq:Eg-VBS-12}
\frac{E_{0}}{J}=-\frac{3}{8}-\frac{3}{32} \lambda^{2}-\frac{3}{256}\lambda^{3}-0.0097656\lambda^{4}-0.0079753\lambda^{5}.
\nonumber
\end{equation}
This leads to the estimation of the ground state energy, $-0.498 J$ for the VBS-12 phase.

Following similar procedures, the ground state energy
of the VBS-36 phase is obtained as
\begin{equation}\label{eq:Eg-VBS-36}
\frac{E_{0}}{J}=-\frac{3}{8}-\frac{1}{16} \lambda^{2}-\frac{5}{256}\lambda^{3}-0.0107422\lambda^{4}-0.0055141\lambda^{5}.
\nonumber
\end{equation}
This leads to the estimation of the ground state energy, $-0.473 J$ for the VBS-36 phase,
which is higher than that of the VBS-12 state. This is consistent with the previous results of the
perturbation theory and the VBS-12 phase is the ground state of the $J_{4}$-depleted
Kagome lattice.

In summary, the VBS-12 phase is lower in energy on the
$J_{4}$-depleted Kagome lattice ($J_{4}=0$)
while the VBS-36 phase has lower energy on the ideal Kagome lattice ($J_{4}=1$).
Based on this result, we expect that there should be an energy level crossing between
these two limits as the magnitude of $J_{4}/J$ is changed from zero to one.
In order to obtain the detailed information about the level crossing, we use
the self-consistent bond operator mean field approach to the two phases for an
arbitrary strength of $J_4/J$ in the next section.

\subsection{\label{sec:BondOperator} Bond Operator Mean Field Theory}

We study the relative stability of the VBS-12 and VBS-36 phases using the
bond operator mean field theory \cite{chubukov,sachdev, Gopalan} for the deformed
Kagome lattice with an arbitrary strength of $J_4/J \le 1$.
In this formulation, the dimer singlet degrees of freedom
are used as natural building blocks and the quantum
corrections coming from the triplet fluctuations
can systematically be investigated \cite{Kotov1,Kotov2}.

\begin{figure}[t]
\centering
\includegraphics[width=8 cm]{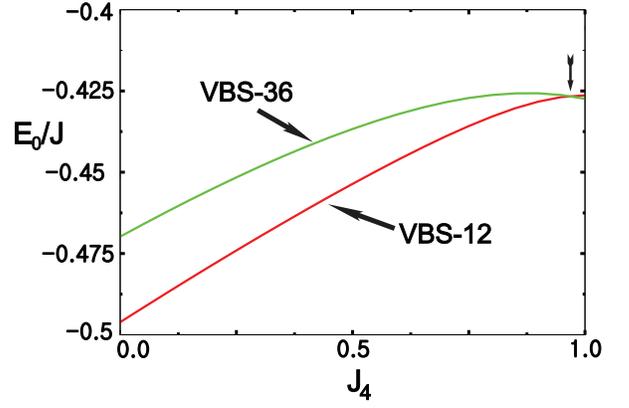}
\caption{(Color online)
Ground state energies of the VBS-12 and the VBS-36 phases
as a function of $J_{4}/J$ ($J$ is set to one) obtained from the bond operator
mean field theory. Note that there is a
level crossing around $J_{4} = 0.97 J$ as indicated by an arrow.
}\label{fig:GroundEnergy}
\end{figure}

Let us consider the two $S=\frac{1}{2}$ spins constituting
a dimer singlet, ${\bf S}_R$
and ${\bf S}_L$. The Hilbert space is spanned by four states
that can be taken as a singlet state, $|s\rangle$, and three
triplet states, $|t_{x}\rangle$, $|t_{y}\rangle$ and
$|t_{z}\rangle$. Then, the singlet and triplet boson operators are
introduced such that each of the above states can be created from
the vacuum $|0\rangle$ as follows:
\begin{align}
|s\rangle  &=s^{\dagger} |0\rangle  =\frac{1}{\sqrt{2}}
(|\uparrow\downarrow\rangle-|\downarrow\uparrow\rangle ),
\nonumber\\
|t_{x}\rangle  &=t_{x}^{\dagger} |0\rangle  =-\frac{1}{\sqrt{2}}
(|\uparrow\uparrow\rangle-|\downarrow\downarrow\rangle ),
\nonumber\\
|t_{y}\rangle  &=t_{y}^{\dagger} |0\rangle  =\frac{i}{\sqrt{2}}
(|\uparrow\uparrow\rangle+|\downarrow\downarrow\rangle ),
\nonumber\\
|t_{z}\rangle  &=t_{z}^{\dagger} |0\rangle  =\frac{1}{\sqrt{2}}
(|\uparrow\downarrow\rangle+|\downarrow\uparrow\rangle ).
\nonumber
\end{align}
To eliminate unphysical states from the enlarged Hilbert space,
the following constraint needs to be imposed on the bond-particle
Hilbert space:
\begin{equation} \label{eq:constraint}
s^{\dagger}s +t_{\alpha}^{\dagger}t_{\alpha} = 1,
\nonumber
\end{equation}
where $\alpha=x,y,$ and $z$, and we adopt the summation convention
for the repeated indices hereafter unless mentioned otherwise.

Constrained by this equation,
the exact expressions for the spin operators can be written in
terms of the bond operators:
\begin{align}\label{eq:bond-ops}
S_{R\alpha}&=\frac{1}{2}(s^{\dag}t_{\alpha} +t_{\alpha}^{\dag}s
-i\varepsilon_{\alpha\beta\gamma}t_{\beta}^{\dag}t_{\gamma}),
\nonumber\\
S_{L\alpha}&=\frac{1}{2}(-s^{\dag}t_{\alpha} -t_{\alpha}^{\dag}s
-i\varepsilon_{\alpha\beta\gamma}t_{\beta}^{\dag}t_{\gamma}),
\nonumber
\end{align}
where $\varepsilon_{\alpha\beta\gamma}$ is the third-rank
totally antisymmetric tensor with $\varepsilon_{xyz}=1$.

We consider the antiferromagnetic Heisenberg model Hamiltonian,
$H = \sum_{\langle ij \rangle } J_{ij}
\textbf{S}_{i} \cdot \textbf{S}_{j}$, where
the exchange couplings of the neighboring spins
are taken to be $J_1=J_2=J_3=J \ge J_4$ (See Fig.~\ref{fig:LatticeStructure}).
Utilizing the bond operator representation of spin
operators, the Hamiltonian can be rewritten solely in terms of bond
particle operators. The hard-core constraint among
bond particle operators is imposed via the Lagrange multiplier
method. The quartic interactions between the triplets in the resulting
Hamiltonian are treated by the self-consistent Hartree-Fock approximation.

The computational procedure for the VBS-36 phase is basically in parallel to
that of the recent work on the 36-site-unit-cell VBS ground state of the
ideal Kagome lattice \cite{Yang}.
Here we present some details of the bond operator theory applied to
the VBS-12 phase on the deformed Kagome lattice.
Since the six dimers within the unit cell of the VBS-12 phase are
equivalent as is obvious from the six-fold rotational symmetry of
the ordering pattern in Fig.\ref{fig:VBSStructure}(a),
the singlet condensate density
$\langle s_{\textbf{i},n} \rangle$ and the chemical potential $\mu_{\textbf{i},n}$
can be set to be $\langle s_{\textbf{i},n} \rangle$ = $\bar{s}$ and
$\mu_{\textbf{i},n}$=$\mu$ in our mean field theory.
Here ${\bf i}$ denotes the
position of the unit cell and $n$ indicates the
dimer index inside the unit cell.
The hard-core constraint on the bond-particle operators is imposed by
adding the following Lagrange multiplier term,
$H_{\mu} = -\sum_{{\bf i},n} \mu (\bar{s}^2 + t^{\dagger}_{{\bf i},n\alpha}t_{{\bf i},n\alpha}-1)$.

The quartic interactions between the triplet particles are decoupled using
the mean field order parameters $P_{\gamma}$ and $Q_{\gamma}$, where
$P_{\gamma} \equiv \langle t^{\dagger}_{{\bf i},n\alpha}t_{{\bf j},m\alpha}  \rangle$
and $Q_{\gamma} \equiv \langle t_{{\bf i},n\alpha}t_{{\bf j},m\alpha}  \rangle$.
We consider four mean-field variables, $P_{\gamma}$ and $Q_{\gamma}$,
labeled by $\gamma$ = 1 or 2. Here $P_{1}$ and $Q_{1}$ denote
the diagonal and off-diagonal triplet correlations between the dimers
residing in the {\it same} unit cell while the other two variables, $P_{2}$ and $Q_{2}$
represent the triplet correlations between neighboring dimers
in {\it different} unit cells. The above four order parameters
$P_{1}$, $Q_{1}$, $P_{2}$, and $Q_{2}$
together with $\bar{s}$ and $\mu$ are determined self-consistently
by solving the coupled saddle point equations. \cite{sachdev,Gopalan}

\begin{figure}[t]
\centering
\includegraphics[width=8 cm]{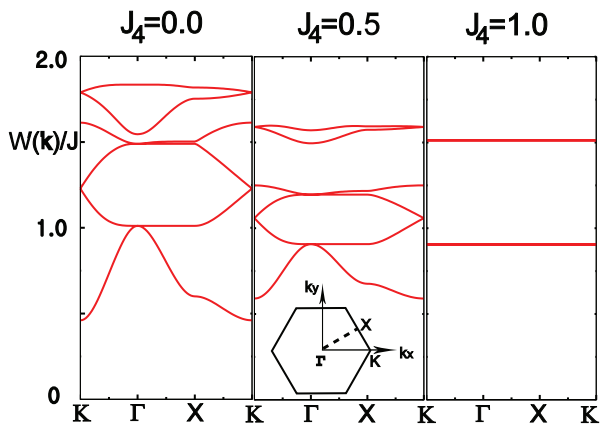}
\caption{(Color online)
Evolution of the triplet dispersion spectra of the VBS-12 phase as
a function of $J_{4}/J$ ($J$ is set to one). Note that the lower flat mode
is fourfold degenerate while the upper one is doubly degenerate
when $J_{4}/J=1$.
} \label{fig:Dispersion}
\end{figure}

In Fig.~\ref{fig:GroundEnergy}, we plot the ground state energies
of the VBS-12 and VBS-36 phases as a function of $0 \le J_{4}/J \le 1$.
When $J_{4}=0$, the ground state energy of the VBS-12 (VBS-36)
from the bond operator mean field theory is $- 0.496 J$ ($- 0.470 J$)
which is close to the results from the series expansion study,
$- 0.498 J$ ($- 0.473 J$). As one increases the magnitude of $J_{4}$,
the ground state energy of the VBS-12 grows more rapidly
than that of the VBS-36 phase and finally an energy level crossing
occurs around $J_{4} = 0.97 J$. Beyond this point the VBS-36 phase
becomes more stable and remains as the ground state up to the ideal
Kagome lattice limit.

The evolution of the triplet excitation spectrum for
the VBS-12 phase shows
interesting behavior as the magnitude of $J_{4}$ changes.
In Fig.~\ref{fig:Dispersion}, we plot the triplet dispersion
spectra when $J_{4} = 0, 0.5$ and 1, respectively. Note that
the triplet dispersion of the VBS-12 phase on the ideal Kagome
lattice ($J_4=J$) shows completely flat spectrum.
Flat dispersion in the momentum space indicates the existence
of localized eigenstates.\cite{Yang} In this case, the three neighboring
dimers around an empty triangle support the localized eigenmodes.
When $J_{4}=J$, the six dimers within a unit cell are all `orthogonal' to
each other in the sense that each spin of a dimer is simultaneously
connected to the two spins of its neighboring dimers as shown in Fig.~\ref{fig:VBSStructure}(a).
In this situation, the triplet particles can fluctuate only through the dimers
of its neighboring unit cell connected via empty triangles, and hence they
are confined within that space.
The three localized triplet modes coming from
the three dimers around an empty triangle
consist of doubly degenerate E modes and a nondegenerate A mode
(in the standard convention of irreducible representations in group theory.\cite{Tinkham}).
Since the unit cell consists of six dimers, each E mode and A mode
is doubly degenerate. In Fig.~\ref{fig:Dispersion}(c),
the lower flat mode has fourfold degeneracy while the upper one
is doubly degenerate.

The fact that all the triplet dispersions are completely flat
is an interesting characteristic of the triplet spectrum of the
VBS-12 phase on the ideal Kagome lattice (note, however, that
the true ground state is the VBS-36 phase on the ideal Kagome lattice).
According to a recent work \cite{Yang},
the triplet spectra of the VBS-36 phase on the ideal Kagome lattice
contain some dispersive modes even though the number of flat modes
is quite large there as well.
In particular, the lowest flat band is degenerate with the second lowest
dispersive band only at the zone center. As discussed in detail in
Ref.\onlinecite{Bergman,Yang}, the degeneracy at a particular momentum point
reflects the existence of noncontractible loop states and is topologically protected.
Therefore, the completely flat spectrum of the VBS-12 phase on the ideal
Kagome lattice means that all the flat modes are topologically trivial.
This results from the fact that the size of every localized mode is not larger than
the size of the unit cell and each localized eigenmode around an empty triangle
is linearly independent. The condition for the emergence of the topologically
nontrivial structure in dispersions is discussed in detail in Ref.\onlinecite{Bergman}.
(See also Ref.\onlinecite{Yang}.)

On the other hand, when the magnitude of $J_{4}$ is reduced,
the degenerate bands split and triplet particles spread over the whole lattice.
In the case of $J_{4} = 0.5 J$ that may directly be relevant to Rb$_{2}$Cu$_{3}$SnF$_{12}$,
one can still see the gap between the lower four bands and the upper two bands.
However, when $J_{4}$ becomes vanishingly small, the gap structure disappears due to
wide bandwidth of the triplet bands.

\section{\label{sec:discussion} Discussion}

\begin{figure}[t]
\centering
\includegraphics[width=8 cm]{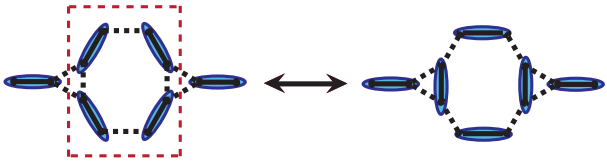}
\caption{(Color online)
A local singlet resonance in the VBS-12 phase
on the J$_{4}$-depleted Kagome lattice. Here the
dotted (red) square represents the smallest flippable loop.
} \label{fig:Loopflip}
\end{figure}

In the case of the VBS phases with a big unit cell, a large number of degenerate
dimer configurations may allow many singlet excitations below a spin gap. For example,
the VBS-36 order on the ideal Kagome lattice leads to the large degeneracy that comes from
the combination of the broken translational symmetry and the local resonance moves
around flippable loops (See Ref.\onlinecite{Nikolic} and Ref.\onlinecite{Huse2}).
Here the flippable loops mean a collection of alternating occupied and unoccupied links;
when the roles of the occupied and unoccupied links
on a flippable loop are interchanged, one hardcore dimer covering
can be continuously transformed to another one. The perfect hexagons
and the pinwheel dimers form the smallest flippable loops in the case
of the 36-site VBS phase on the ideal Kagome lattice.\cite{Nikolic}
The number of singlet excitations arising from such considerations is still smaller
than what was observed in the exact diagonalization studies \cite{Sindzingre, Lecheminant, Misguish1, Misguish2}
of small system sizes (up to 36 sites) for the spin-1/2 antiferromagnetic Heisenberg model on the ideal
Kagome lattice. It was, however, argued that the triplets can scatter to form additional
singlet bound states in finite size systems and this may explain the exact diagonalization results \cite{Huse2}.

On the other hand, the VBS-12 phase on the $J_{4}$-depleted Kagome lattice
does not break any translational symmetry and the local resonances from loop flips
are the only sources of low energy singlet excitations.
In Fig.\ref{fig:Loopflip}, we show the smallest flippable loop
consisting of the four occupied and four unoccupied links, and its loop flip
in the VBS-12 phase. Note that there is a change in the number
of the {\it empty} and {\it filled} triangles after the loop flips.
In fact, a single loop flip increases the number of the {\it filled} triangles by two.
According to Eq.(\ref{eq:E-secondorder}) and Eq.(\ref{eq:E-thirdorder}) in the perturbation theory,
this requires the energy cost of
$-4\Delta \epsilon^{(2)} + 2\Delta \epsilon^{(3)}_{a} = \frac{21}{64} J$,
which is much larger than the measured spin gap $\sim 20K$
if the experimentally estimated $J \sim 200K$ is used.
Since the number of the {\it filled} triangles involved in a loop flip is
proportional to the size of the flippable loop, the singlet excitations
coming from a larger loop flip require even higher energy cost.
Therefore, the VBS-12 phase is quite stable against local singlet fluctuations
and it seems difficult to produce ``the singlet wave" whose existence is
speculated in a recent experimental paper. \cite{Exp}
We expect the lowest singlet excitation would be well
above the spin gap, which can be contrasted with many
singlet excitations below a spin gap in the case of the ideal Kagome lattice.

In summary, slight lattice distortions can completely change
the nature of the ground state of the quantum antiferromagnet
on the Kagome lattice and this leads to the abrupt
modification of the elementary excitation spectra in both the
triplet and singlet channels. Even though $J_1$, $J_2$, and $J_3$ are
not exactly the same in Rb$_{2}$Cu$_{3}$SnF$_{12}$, the magnitudes of
these exchange couplings are very similar and satisfy
$J_1 > J_2 > J_3 \gg J_4$. Interestingly, the VBS-12 phase studied
in this work corresponds to the dimer covering that resides only on
the $J_1$ links. This suggests that the VBS-12 phase on the $J_1$ links
is stable all the way to the limit where $J_2$ and $J_3$ become the same
as $J_1$. Thus, we expect that our results are directly applicable to
Rb$_{2}$Cu$_{3}$SnF$_{12}$.
In particular, our predictions on the triplet
and singlet spectra can be tested in the future neutron scattering
and specific heat experiments on Rb$_{2}$Cu$_{3}$SnF$_{12}$.

\acknowledgments

We thank Tyler Dodds, Michael J. Lawler, and Erik S. S{\o}rensen for helpful discussions.
This work was supported by the NSERC of Canada, the Canada Research Chair, and
the Canadian Institute for Advanced Research.





\begin{thebibliography}{99}

\bibitem{Misguich_book} G. Misguich and C. Lhuillier,
``Two-dimensional Quantum Antiferromagnets", published in {\it
Frustrated Spin Systems}, edited by H. T. Diep, pp. 229-306 (World
Scientific Publishing, 2004).

\bibitem{sachdev_kagome} S. Sachdev, Phys. Rev. B {\bf 45}, 12377 (1992).

\bibitem{Fa_PSG} F. Wang and A. Vishwanath, Phys. Rev. B {\bf 74}, 174423 (2006).

\bibitem{Hastings} M. B. Hastings, Phys. Rev. B {\bf 63}, 014413 (2000).

\bibitem{Ran} Y. Ran, M. Hermele, P. A. Lee, and X. G. Wen, Phys. Rev. Lett.
{\bf 98}, 117205 (2007).

\bibitem{Ryu} S. Ryu, O. I. Motrunich, J. Alicea, and M. P. A. Fisher, Phys. Rev. B
{\bf 75}, 184406 (2007).

\bibitem{Marston} J. B. Marston and C. Zeng, J. Appl. Phys. {\bf
69}, 5962 (1991).

\bibitem{Nikolic} P. Nikolic and T. Senthil, Phys. Rev. B
{\bf 68}, 214415 (2003).

\bibitem{Huse} R. R. P. Singh and D. A. Huse, Phys. Rev. B {\bf 76}, 180407(R)
(2007).

\bibitem{Huse2} R. R. P. Singh and D. A. Huse, Phys. Rev. B {\bf 77}, 144415 (2008).

\bibitem{Fa} F. Wang, A. Vishwanath, and Y. B. Kim, Phys. Rev. B
{\bf 76}, 094421 (2007).

\bibitem{Lawler} M. J. Lawler, L. Fritz, Y. B. Kim, and S. Sachdev, Phys. Rev. Lett
{\bf 100}, 187201 (2007).

\bibitem{Exp_mag1} N. Rogado, M. K. Haas, G. Lawes, D. A. Huse, A. P. Ramirez,
and R. J. Cava, J. Phys. Condens. Matter, {\bf 15}, 907 (2003).

\bibitem{Exp_mag2} Y. Yamabe, T. Ono, T. Suto, and H. Tanaka
, J. Phys. Condens. Matter, {\bf 19}, 145253 (2007).


\bibitem{Exp_vesign} Y. Okamoto, H. Yoshida, Z. Hiroi, arxiv:0901.2237.

\bibitem{Exp_herbert1} M. P. Shores, E. A. Nytko, B. M. Bartlett,
and D. G. Nocera, J. Am. Chem. Soc. {\bf 127}, 13462
(2005).

\bibitem{Exp_herbert2} J. S. Helton, K. Matan, M. P. Shores, E. A. Nytko,
B. M. Bartlett, Y. Yoshida, Y. Takano, A. Suslov, Y. Qiu, J.-H.
Chung, D. G. Nocera, and Y. S. Lee, Phys. Rev. Lett. {\bf 98},
107204 (2007).


\bibitem{Exp_herbert3} P. Mendels, F. Bert, M. A. de Vries, A. Olariu, A. Harrison, F.
Duc, J. C. Trombe, J. S. Lord, A. Amato, and C. Baines , Phys.
Rev. Lett. {\bf 98}, 077204 (2007).

\bibitem{Exp_volbo} Z. Hiroi, M. Hanawa, N. Kobayashi,
H. Takagi, Y. Kato, and M. Takigawa, J.Phys.Soc.Jpn.  {\bf 70}, 3377
(2001).

\bibitem{Exp} K. Morita, M. Yano, T. Ono, H. Tanaka, K. Fugii,
H. Uekusa, Y. Narumi, and K. Kindo, J.Phys.Soc.Jpn.  {\bf 77}, 043707
(2008).








\bibitem{perturbation} T.Barnes, J.Riera, and D. A. Tennant, cond-mat/9801224.


\bibitem{SeriesBook} J. Oitmaa, C. Hamer and W-H. Zheng, Series
Expansion Methods for Strongly Interacting Lattice Models
(Cambridge University Press, Cambridge, 2006).

\bibitem{SeriesBook2} C.Domb and M.S.Green, Phase Transitions and
Critical Phenomena (Academic Press, New York, 1974), Vol. 3.

\bibitem{Rigol1} M. Rigol, T. Bryant and R. R. P. Singh , Phys. Rev. E {\bf 75}, 061118
(2007).

\bibitem{Rigol2} M. Rigol, T. Bryant and R. R. P. Singh , Phys. Rev. Lett {\bf 97}, 187202
(2006).
















\bibitem{chubukov} A. V. Chubukov, JETP Lett. {\bf 50}, 129 (1989).

\bibitem{sachdev} S. Sachdev and R. N. Bhatt, Phys. Rev. B {\bf 41},
9323 (1990).

\bibitem{Gopalan} S. Gopalan, T. M. Rice and M. Sigrist, Phys. Rev. B
{\bf 49}, 8901 (1994).

\bibitem{Kotov1} V. N. Kotov, O. Sushkov, Zheng Weihong, and J. Oitmaa, Phys. Rev.
Lett. {\bf 80}, 5790 (1998).

\bibitem{Kotov2} O. P. Sushkov and V. N. Kotov, Phys. Rev. Lett. {\bf 81},
1941 (1998).

















\bibitem{Yang} B.-J. Yang, Y. B. Kim, J. Yu, K. Park, Phys. Rev. B {\bf 77}, 224424 (2008).

\bibitem{Tinkham}
M. Tinkham, Group Theory and Quantum Mechancies, (McGraw-Hill, New York, 1964).

\bibitem{Bergman} D. L. Bergman, C. Wu, and L. Balents,
Phys. Rev. B {\bf 78}, 125104 (2008).


\bibitem{Lecheminant} P. Lecheminant, B. Bernu, C. Lhuillier,
L. Pierre,and P. Sindzingre, Phys. Rev. B {\bf
56}, 2521 (1997).

\bibitem{Sindzingre} Ch. Waldtmann, H.-U. Everts, B. Bernu, C. Lhuillier,
P. Sindzingre, P. Lecheminant, and L. Pierre, Eur. Phys. J. B {\bf
2}, 501 (1998).

\bibitem{Misguish1} G. Misguish and B. Bernu, Phys. Rev. B {\bf
71}, 014417 (2005).

\bibitem{Misguish2} G. Misguish and P. Sindzingre,
J.Phys.: Condens. Matter {\bf
19}, 145202 (2007).


\end{thebibliography}
\end{document}